\documentclass[CRMECA,Unicode,manuscript,biblatex]{cedram}
\usepackage{bm}
\usepackage[dvipsnames]{xcolor}
\usepackage{siunitx}
\usepackage{tikz}
\newcommand{\tikzcircle}[2][red,fill=red]{\tikz[baseline=-0.8ex]\draw[#1,radius=#2] (0,0) circle ;}%
\definecolor{Reds1}{rgb}{0.573,0.208,0.118}
\definecolor{Reds2}{rgb}{0.827,0.467,0.314}
\definecolor{Reds3}{rgb}{0.925,0.698,0.49}
\definecolor{Reds4}{rgb}{0.984,0.89,0.761}
\definecolor{Blues1}{rgb}{0.031,0.271,0.58}
\definecolor{Blues2}{rgb}{0.129,0.443,0.71}
\definecolor{Blues3}{rgb}{0.259,0.573,0.776}
\definecolor{Blues4}{rgb}{0.42,0.682,0.839}
\definecolor{Blues5}{rgb}{0.62,0.792,0.882}
\definecolor{Blues6}{rgb}{0.776,0.859,0.937}
\newcommand{\revP}[1]{\textcolor{Black}{#1}}

\title{Wave propagation in a model artery}

\author{\firstname{Pierre} \lastname{Chantelot} \CDRorcid{0000-0003-1342-2539} \IsCorresp}
\address{Institut Langevin, ESPCI Paris, Université PSL, CNRS, 75005 Paris}
\email[P. Chantelot]{p.chantelot@gmail.com}
\author{\firstname{Alexandre} \lastname{Delory}\CDRorcid{0000-0002-7623-5856}}
\address{ENS de Lyon, CNRS, LPENSL, UMR 5672, Lyon cedex 07, France}

\author{\firstname{Claire} \lastname{Prada}\CDRorcid{0000-0002-2500-1099}}
\addressSameAs{1}{Institut Langevin, ESPCI Paris, Université PSL, CNRS, 75005 Paris}

\author{\firstname{Fabrice} \lastname{Lemoult}\CDRorcid{0000-0001-9757-0760}\IsCorresp}
\addressSameAs{1}{Institut Langevin, ESPCI Paris, Université PSL, CNRS, 75005 Paris}
\email[F. Lemoult]{fabrice.lemoult@espci.psl.eu}
\thanks{This work has been partially supported by the Simons Foundation/Collaboration on Symmetry-Driven Extreme Wave Phenomena and received support under the program ``Investissements d'Avenir" by the French Government.} 
\ESM{The scripts to compute dispersion relations and data underlying this work are available on Zenodo (DOI: \href{https://zenodo.org/records/16037558}{10.5281/zenodo.16037557}) and \href{https://github.com/pchantelot/Wave_propagation_in_a_model_artery_data}{GitHub}, respectively.}
\keywords{Soft matter, hyperelasticity, viscoelasticity, pulse wave}

\begin{abstract} 
Fluid filled pipes are ubiquitous in both man-made constructions and living organisms. In the latter, biological pipes, such as arteries, have unique properties as their walls are made of soft, incompressible, highly deformable materials.
In this article, we experimentally investigate wave propagation in a model artery: an elastomer strip coupled to a rigid water channel.
We measure out-of-plane waves using synthetic Schlieren imaging, and evidence a single dispersive mode which resembles the pulse wave excited by the heartbeat.
By imposing an hydrostatic pressure difference, we reveal the strong influence of pre-stress on the dispersion of this wave.
Using a model based on the acoustoelastic theory accounting for the material rheology and for the large static deformation of the strip, we demonstrate that the imposed pressure affects wave propagation through an interplay between stretching, orthogonal to the propagation direction, and curvature-induced rigidity.
We finally highlight the relevance of our results in the biological setting, by discussing the determination of the arterial wall's material properties from pulse wave velocity measurements in the presence of pre-stress.
\end{abstract}

\begin{document}

\maketitle

% Example of section
\section{Introduction}
The mechanical properties of cells, tissues, and organs are intimately tied to their physiological functions \cite{levental2007}.
As a result, pathologies are often associated with alterations of the mechanical behavior of tissues: for example, alveolar lung disease is linked to a reduction of tissue shear modulus \cite{yuan2000}, and tumor cells are significantly stiffer than healthy ones \cite{kumar2009}.
Developing tools to probe the material properties of biological tissues \emph{in vivo} is thus crucial to improve diagnosis.
In the case of medical imaging, the development of shear wave elastography has shown that monitoring the propagation of elastic waves allows to quantitatively assess material properties \cite{bercoff2004, sigrist2017}.
More generally, relating the velocity of elastic waves to the medium characteristics is relevant to a broad scope of applications ranging from non-destructive testing in industry \cite{rogers1995,bochud2018} to geophysics imaging \cite{shapiro2005}.

Here, we focus on the link between the propagation of elastic waves in arteries and the stiffness of the arterial wall, a widely used marker of cardiovascular risk \cite{ben2014,safar2018}. 
An elastic wave is naturally excited as the heart beats and travels along the arterial tree, as first described by Young \cite{young1809}.
Locally measuring the propagation of this wave, called the pulse wave, and estimating the corresponding tissue elasticity constitutes the medical gold standard \cite{engelen2015,bossuyt2015}.
The tissue stiffness is usually inferred from the Moens-Korteweg formula \cite{moens1878,korteweg1878}, assuming a breathing motion and that the wave speed is set by an interplay between the linear elastic response of the wall and blood inertia.
This approach captures the essential physics of pulse wave propagation, but it fails to account for several clinical observations.
It overlooks the complex structure of the arterial wall tissue, consisting of three distinct layers, which gives arteries anisotropic and non-linear material properties \cite{amabili2021}.
It does not recognize that multiple modes are excited by the heartbeat, as demonstrated by the recent \emph{in vivo} observation of an arterial flexion wave \cite{laloy2023,gregoire2024}.
It ignores that the pulse wave velocity is altered by pre-stress: variations of the mean blood pressure affect the pulse wave velocity \cite{histand1973,marais2018,baranger2023}, and waves generated during elastography measurements are sensitive to the large transient deformations occurring during the cardiac cycle \cite{couade2010,li2017}.
The pulse wave thus probes the incremental stiffness of the arterial wall \cite{pedley1980}.
Additionally, elastography data highlight that wave guiding and viscoelastic material properties also influence wave propagation in arteries \cite{couade2010,bernal2011,maksuti2016,roy2021}.

Acknowledging these limitations, a rich literature builds upon the seminal work of Moens and Korteweg, addressing the finite deformations of tubes \cite{haughton1979a,haughton1979b}, the material non-linearity of the arterial wall \cite{holzapfel2010,ogden2016}, the influence of wave guiding and viscoelasticity \cite{couade2010,roy2021}, and the role of pre-stress on wave propagation in fluid-loaded plates and tubes \cite{kuiken1984, demiray1996, fu2010, li2017}.
In this article, we aim to integrate all these aspects into a comprehensive model that can be quantitatively compared with experimental data, ultimately allowing an accurate assessment of the arterial wall's material properties.

We first describe the guided waves that propagate in soft elastic tubes filled and surrounded by water. 
We choose to focus on the lowest order modes that are the most physiologically relevant, and provide analytical long-wavelength approximations for the propagation speed of the breathing and flexion modes observed \emph{in vivo} (section \ref{sec:pipewaves}).  
Next, we introduce our experiment which consists in a square rigid conduit closed by an elastomer wall.
This minimal artery phantom allows to successfully isolate a single mode resembling the breathing mode associated to the pulse wave. 
We impose a static fluid pressure, creating pre-stress, and highlight its influence on the dispersion relation of the breathing mode (section \ref{sec:arteryexperiment}).
In section \ref{sec:prestress}, we perform dedicated experiments, conducted in the absence of fluid coupling, to evidence how transversal stretching and curvature-induced rigidity combine through the deflection of the soft wall. 
We quantify the influence of pre-stress in a semi-analytical model, based on the acoustoelastic theory, that accounts for wave guiding, viscoelasticity, geometrical and material non-linearities.
In section \ref{sec:fluidcoupling}, we tackle fluid coupling in the short and long-wavelength limits relevant to shear wave elastography and natural pulse wave propagation, respectively.
Finally, we discuss the relevance of our results in inferring material properties from wave velocity measurements, with the aim of going beyond the determination of incremental elastic moduli (section \ref{sec:discussion}).
\section{Axial waves in a soft pipe}
\label{sec:pipewaves}
\begin{figure}
    \centering
    \includegraphics[width = \textwidth]{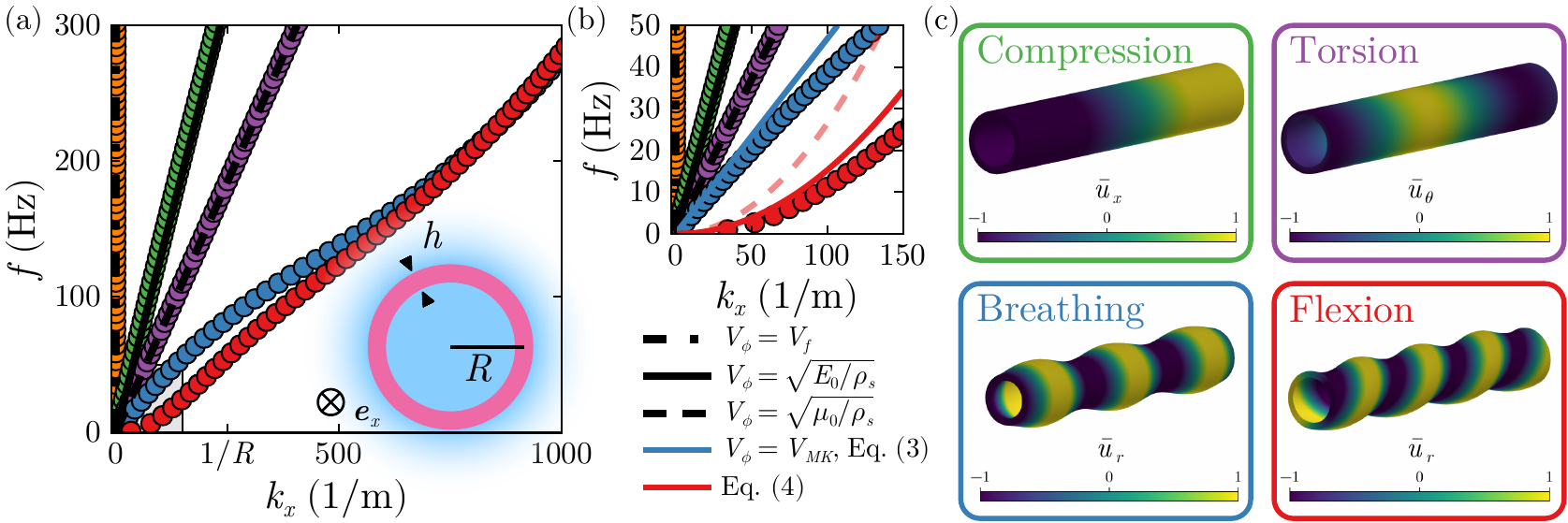}
    \caption{(a) Dispersion relation of the axial waves propagating along a water filled nearly incompressible elastic pipe ($\rho_s = \qty{1070}{\kilogram\per\cubic\meter}$, $\mu_0 = \rho_s V_T^2 = \qty{23}{\kilo\pascal}$) with radius $R = \qty{4}{\mm}$ and thickness $h = \qty{1}{\mm}$ surrounded by an infinite water medium. 
    We represent only the six lowest order fluid (orange), compression (green), torsion (purple), breathing (blue), and flexion (red) modes.  
    (b) Zoom of (a) in the small wavenumber limit, $k_x R \ll 1$ (greyed region), where one dimensional models successfully capture the dispersion of axial waves. 
    (c) Displacement fields at 100 Hz for the solid zero order modes identified in (a), where the color code corresponds to the main component of the normalized displacement field for each mode.}
    \label{fig1}
\end{figure}
We consider a water-filled isotropic elastic tube with radius $R$, thickness $h$, density $\rho_s$, shear and longitudinal wave velocities, $V_T$ and $V_L$ respectively. 
The tube is immersed in an infinite water medium (density $\rho_f = \qty{1000}{\kilogram\per\meter\cubed}$ and speed of sound $V_f = \qty{1500}{\metre\per\second}$), as sketched in the inset of figure \ref{fig1}(a).
\revP{This configuration approximates that of a blood vessel embedded in its surrounding as the density and speed of sound of blood and biological tissues are close to that of water.}
Given this geometry, we focus on computing the dispersion relation of guided waves propagating along the direction of the pipe.
Our aims are (i) to relate these results to \emph{in vivo} observations of wave propagation in blood vessels, and (ii) to assess the ability of long-wavelength models to predict the relation between the tube wall stiffness and the pulse wave velocity. 

\subsection{Three-dimensional model}
The displacement field $\boldsymbol{u}$ in the inner fluid domain and in the tube wall satisfies the three-dimensional elastodynamics equation,
\begin{equation} 
    \boldsymbol{\nabla} \cdot (\boldsymbol{C}_i : \boldsymbol{\nabla} \boldsymbol{u}) = \rho_i \ddot{\boldsymbol{u}},
    \label{eq:elastodynamics}
\end{equation}
where $\boldsymbol{C}$ is the fourth-order elasticity tensor, and the subscript $i$ denotes fluid ($f$) or solid ($s$) material properties (note that we model the fluid as an isotropic elastic solid with $V_L = V_f$ and $V_T / V_f \approx 10^{-9}$).
The inner fluid and the elastic wall are coupled at the inner tube surface ($r = R - h/2$) by imposing the continuity of both the normal traction and the normal displacement. 
The presence of the surrounding fluid is taken into account in the boundary condition for the normal stress at the outer surface of the tube, 
\begin{equation}
    \boldsymbol{e}_r\cdot(\boldsymbol{C} : \boldsymbol{\nabla} \boldsymbol{u})|_{r=R+h/2} =  \frac{\rho_f\omega^2}{\gamma}\frac{H_m\left(\gamma r\right)}{H'_m\left(\gamma r\right)}  u_r \boldsymbol{e}_r,
    \label{eq:cylinderbc}
\end{equation}
where $H_m$ is the Hankel function of the first kind of order $m$, and $\gamma = \sqrt{k_f^2 - k_x^2}$ with $k_f$ the wavenumber in the fluid, and $k_x$ the axial projection of the wavenumber $\boldsymbol{k}$ in the pipe \cite{junger1952,gravenkamp2014}.
After inserting the displacement ansatz, $\boldsymbol{u} = \boldsymbol{u}(k_x,r,m,\omega)e^{\mathrm{i}(k_x x  + m\theta - \omega t)}$, the dispersion relation $k_x(\omega)$ is given by non-trivial solutions of the system of equations (\ref{eq:elastodynamics}--\ref{eq:cylinderbc}).
Solutions to similar systems, where equation \eqref{eq:elastodynamics} is replaced by equations of motion derived from thin or thick shell theory, have been obtained by finding the zeroes of a transcendental equation \cite{bleich1954,lin1956,warburton1961}.
Here, we use a semi-analytical approach that consists in discretizing the problem in the radial direction using the spectral collocation method \cite{adamou2004,shen2011,kiefer2022}.
The dispersion relation is obtained as the solution $(k_x, \boldsymbol{u})$ to an algebraic eigenvalue problem which incorporates an approximate boundary condition at the tube outer surface. 
This solution is then iteratively refined to fulfill the exact boundary condition given in equation \eqref{eq:cylinderbc} \cite{gravenkamp2014}.
We emphasize that this approach enables us to find complex wavenumbers $k_x$ while keeping $\omega$ real, which is key to modeling viscoelastic media.
Our implementation, adapted from \cite{kiefer2024,gravenkamp2014}, is detailed and available in \cite{chantelot2025}.

\subsection{Dispersion relation}
For the sake of concreteness, we now consider a tube with dimensions similar to that of the carotid artery ($R = \qty{4}{\mm}$ and $h = \qty{1}{\mm}$), and with material properties characteristic of a soft elastomer ($V_T = \qty{4.6}{\meter\per\second}$ and $V_L = \qty{1000}{\meter\per\second}$).
The dispersion relations computed for this set of parameters, considering only the lowest order modes for $m = 0$ and $m = \pm 1$, are shown in figure \ref{fig1}(a). 
Six modes propagate in the low-frequency regime: the fluid mode (orange), the compression mode (green dots), the torsion mode (purple dots), the breathing mode (blue dots), and two degenerated flexion modes (red dots).
The fluid mode, which is not associated to wall motion, as well as the compression and torsion modes, whose displacement fields are predominantly polarized in the axial and azimuthal direction respectively (figure \ref{fig1}c), are barely observable \emph{in vivo}.
Therefore, we focus on the breathing and flexion modes, which have both been observed \emph{in vivo} \cite{laloy2023}, thanks to the radial nature of their displacement field (figure \ref{fig1}c).
When $k_xR \gg 1$, these two modes merge and, in the frequency range relevant to shear wave elastography that is typically from $\qty{100}{\hertz}$ to $\qty{1000}{\hertz}$, both modes are dispersive indicating that waveguiding must be taken into account when analyzing such experiments \cite{couade2010,bernal2011}.
In the long-wavelength limit ($k_xR \ll 1$), the two modes display markedly different behaviors (figure \ref{fig1}b), that we now model and link to the arterial wall material properties. 

\subsection{Long-wavelength approximations for the breathing and flexion modes}
In the long-wavelength limit, the breathing mode, classically associated to the pulse wave, propagates at a constant velocity.
This wave speed can be related to the material properties using the Moens-Korteweg formula \cite{moens1878,korteweg1878}.
We recall here the main steps of the derivation of this formula to evidence the physics of pulse wave propagation.
Using mass and momentum conservation in the fluid contained inside the tube, the speed of a pressure disturbance in the fluid can be expressed as
$V_b = \sqrt{A/(\rho_f \partial A / \partial P)}$, where $A$ is the area of the fluid section, and $P$ the pressure \cite{bramwell1922}.
The response of the tube wall to an internal pressure increase $\mathrm{d}P$ can be derived by expressing the hoop stress using Laplace law for $h \ll R$, and using Hooke's law assuming plane membrane stress in the wall, giving $\mathrm{d}P R/h = E_0 \mathrm{d}R / R$, where $E_0$ is the Young modulus.
We recall that for nearly incompressible materials $E_0 \approx 3\mu_0$.
Combining the two above relations yields the Moens-Korteweg formula,
\begin{equation}
    V_{M\!K} = \sqrt{\frac{E_0 h}{2\rho_f R}},
    \label{eq:MK}
\end{equation}
which evidences that the breathing mode involves both the elasticity of the tube wall and the inertia of the fluid, that is much larger than that of the solid for thin tubes.
It is also worth noticing that, at first-order, the breathing mode velocity is not affected by the surrounding fluid, as evidenced here by the good agreement between the simulated dispersion and equation \eqref{eq:MK} (solid blue line in figure \ref{fig1}b).

In contrast, the flexion mode is dispersive for $k_x R \ll 1$, with a quadratic evolution $\omega \propto k_x^2$, characteristic of bending.
In the case of a fluid-filled tube, this behavior can be captured by a one-dimensional model \cite{chen1974,gregoire2024}.
The dynamics in the long-wavelength limit are described by an Euler-Bernoulli beam equation, $E_0 I \partial^4 u_r / \partial x^4  + \rho_s A \ddot{u}_r = 0 $, where $I$ is the second moment of area and $A$ is the beam cross-section \cite{doyle2021}. 
After inserting the displacement ansatz, we get the dispersion relation $\omega = \sqrt{E_0I/(\rho_sA)} k_x^2$.
This dispersion relation can be modified to take into account the presence of the inner fluid by recognizing that, similarly as for the breathing mode, the inertia of the fluid ($\rho_f \pi R^2$) dominates when $h \ll R$, yielding $\omega = \sqrt{E_0 R h/\rho_f} k_x^2$ \cite{chen1974,gregoire2024}.
Unlike for the breathing mode, this prediction (light red dashed line in figure \ref{fig1}b) does not satisfactorily describe the computed dispersion relation, highlighting the influence of the surrounding fluid.
We incorporate this effect in the one-dimensional model by adding the load from the outer fluid to the beam equation.
This load is the azimuthal integral of the pressure, given by equation \eqref{eq:cylinderbc} with $m = 1$, on the outer tube surface.
We further simplify equation \eqref{eq:cylinderbc} upon noticing that the flexural wave velocity is much slower than that of sound waves in water, indicating that only evanescent waves exist in the outer fluid, \emph{i.e.} $\gamma \approx \mathrm{i} k_x$.
Finally, we obtain the long-wavelength dispersion relation by performing asymptotic expansions of the Hankel functions,
%\begin{equation}
%    \omega = \sqrt{\frac{E_s R e}{\rho_f \left(1 + \frac{2H_1\left(ik_x(R+e/2)\right)}{ik_xRH'_1\left(ik_x(R+e/2)\right)}\right)}} k_x^2,
%    \label{eq:flex}
%\end{equation}
\begin{equation}
    \omega = \sqrt{\frac{E_0 R h}{3\rho_f}} k_x^2,
    \label{eq:flex}
\end{equation}
that we represent with a red solid line in figure \ref{fig1}(b), and that convincingly fits our data for $k_xR \ll 1$.
We point out that the influence of the surrounding fluid appears as an added mass effect here, multiplying by a factor three the mass of fluid set in motion compared to the case with inner fluid only.

Having identified the different modes of wave propagation in arteries, and established the link between the wave velocity and material properties in the long-wavelength limit for the modes observed $\emph{in vivo}$, we perform experiments with the goal to reproduce the physics of the breathing mode.

\section{A minimal artery phantom}
\label{sec:arteryexperiment}

\begin{figure}
    \centering
    \includegraphics[width = \textwidth]{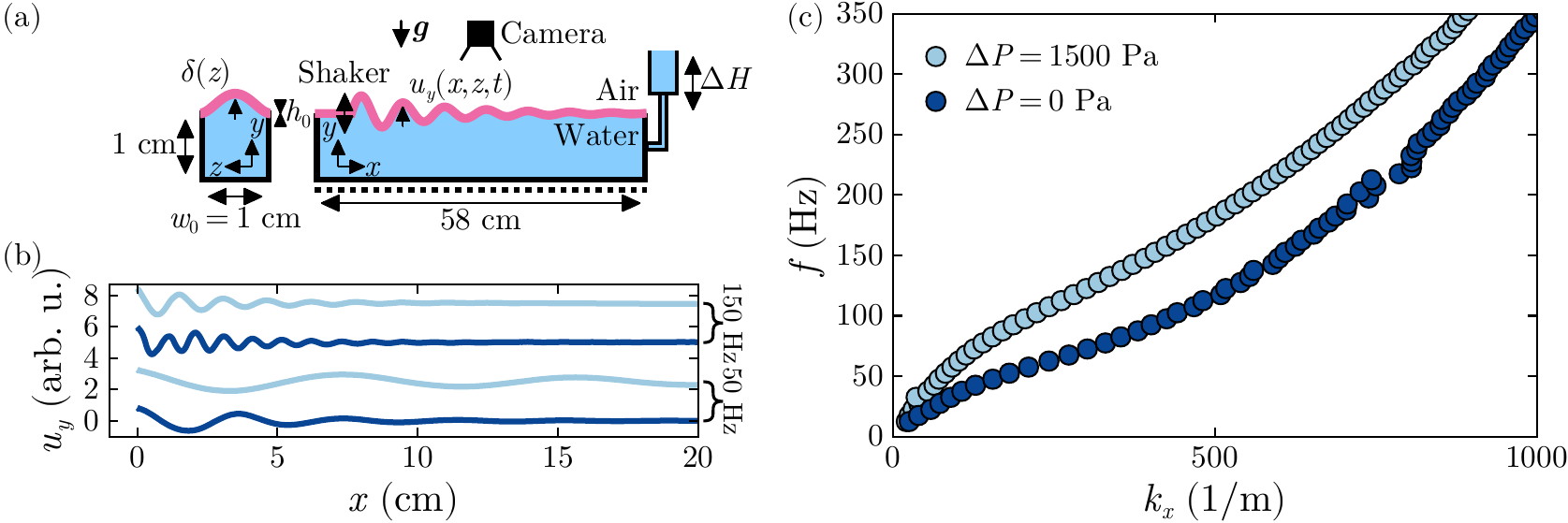}
    \caption{(a) A soft compartment consisting of an elastomer strip ($h_0 =  \qty{1}{\milli\metre}$, $w_0 = \qty{1}{\centi\metre}$) clamped to a rectangular water channel ($\qty{58}{\centi\meter} \times \qty{1}{\centi\meter} \times \qty{1}{\centi\meter}$) is subjected to a pressure difference $\Delta P = \rho_f g \Delta H$, that induces a large deflection of the soft wall $\delta(z)$.
    We generate waves in the strip using a shaker, and measure the out-of-plane displacement field, $u_y(x,z,t)$, associated to wave propagation using Synthetic Schlieren imaging. 
    (b) Out-of-plane displacement field averaged along the z direction for two frequencies: $f = \qty{50}{\hertz}$ and $f = \qty{150}{\hertz}$; and two imposed pressure differences: $\Delta P = \qty{0}{\pascal}$ and $\Delta P = \qty{1500}{\pascal}$.
    (c) Dispersion relations of the zero order mode for $\Delta P = \qty{0}{\pascal}$ and $\Delta P = \qty{1500}{\pascal}$. The mode dispersion closely resembles that of the breathing mode in a fluid-filled tube and is highly sensitive to $\Delta P$.}
    \label{fig2}
\end{figure}

We design an experimental platform that allows to monitor breathing waves and investigate the parameters that influence their propagation properties, while being free of the compression, torsion and flexion modes that propagate along a tube.
\subsection{Experimental setup and procedure}
We build a deformable conduit by clamping a soft elastomer strip with thickness $h_0 = \qty{770}{\micro\meter}$, to the top of a square section, rigid, open Plexiglas channel with dimensions $\qty{58}{\cm} \times \qty{1}{\cm} \times \qty{1}{\cm}$ (figure \ref{fig2}a).
We choose to make the deformable wall from a silicone elastomer, Ecoflex OO-30 (Smooth On), as it shares the low shear modulus, on the order of $\qty{10}{\kilo\pascal}$, and the incompressible nature of biological tissues.
Ecoflex OO-30 exhibits an hyperelastic behavior that we describe using the compressible Mooney-Rivlin model \cite{mooney1940,rivlin1948} which involves two material parameters $\mu_0  = \qty{23}{\kilo\pascal}$ and $\alpha = 0.29$.
These values were taken from tensile tests conducted in a previous work \cite{delory2024}.
Having discussed the response of this elastomer to static strain, we focus on its viscoelastic behavior.
The shear modulus $\mu_s$ is both complex and frequency dependent, and its evolution is well captured by a fractional Kelvin-Voigt model \cite{rolley2019,kearney2015},  $\mu_s = \mu_0\left(1+ \left(\mathrm{i} \omega \tau\right)^n \right)$,
%\begin{equation}
%    \mu_s = \mu_0\left(1+ \left(\mathrm{i} \omega \tau\right)^n \right),
%    \label{KelvinVoigt}
%\end{equation}
where $\tau = \qty{330}{\micro\second}$ and $n = 0.32$ were determined from oscillatory rheological measurements \cite{lanoy2020,delory2024}.
The joint effects of pre-stress \cite{ogden1997,destrade2007} and viscoelasticity \cite{delory2023} are implemented in a modified elasticity tensor $\boldsymbol{C^0}$ which describes the elastomer response for incremental motions superposed on a large static deformation, as detailed in appendix \ref{app:A}. 

We fill the deformable conduit with water and connect it to a water column.
Its height can be adjusted to impose a static pressure difference $\Delta P = \rho_f g \Delta H$ (where $g = \qty{9.81}{\meter\per\second\squared}$ is the gravitational acceleration) between the channel and the ambient.
This pressure difference, that we vary between $\qty{0}{\pascal}$ and $\qty{2900}{\pascal}$, creates a finite deflection of the deformable wall's midline, $\delta(z)$.

After this static pre-stress is applied, a shaker drives a point-like source in the $\boldsymbol{e}_y$ direction, creating small amplitude waves.
The source generates a chirp signal with an instantaneous frequency varying from $\qty{1}{\hertz}$ to $\qty{400}{\hertz}$ in one second, and a CCD camera records images from the top at a frame rate of $\qty{800}{\hertz}$.
The out-of-plane incremental displacement of the soft strip, $u_y(x,z,t)$, induces by refraction a deformation of the pattern placed below the conduit, allowing the use of a synthetic Schlieren imaging method \cite{wildeman2018} to retrieve the displacement field.
As the usual checkerboard pattern would suffer from a large deformation caused by the static deflection of the elastomer, we use a line pattern oriented in the $\boldsymbol{e}_z$ direction.

\subsection{Dispersion relation}
From these measurements, we first compute the monochromatic displacement field $u_y(x,z,f)$ by performing a Fourier transform in time for two imposed pressures $\Delta P = \qty{0}{\pascal}$ and $\Delta P = \qty{1500}{\pascal}$.
The displacement average along the $\boldsymbol{e}_z$ direction is represented in figure \ref{fig2}b for $f = \qty{50}{\hertz}$ and $f = \qty{150}{\hertz}$.
We observe the propagation of a single mode, corresponding to flexural motion of the strip.
These displacement profiles already show the influence of pre-stress. 
For both frequencies, the wavelength and the attenuation distance increase with $\Delta P$. 
This highlights the interplay between the strain on the strip, induced by the hydrostatic pressure, and the viscoelasticity.

We then perform a spatial Fourier transform. 
By extracting maxima for each frequency in the $(k_x, f)$ plane, we retrieve the dispersion relation of the guided axial waves.
We show the result of this analysis for $\Delta P = \qty{0}{\pascal}$ (dark blue) and $\Delta P = \qty{1500}{\pascal}$ (light blue) in figure \ref{fig2}(c).
The dispersion of axial waves is similar to that of the breathing mode in a fluid-filled tube (figure \ref{fig1}a), with a linear behavior at low frequency, followed by an inflection point and a parabolic shape as $f$ increases.
This indicates that we have designed a model experiment that qualitatively reproduces the breathing mode dispersion, and is easier to study as we eliminated the compression, torsion and flexion modes.
Figure \ref{fig2}(c) also highlights the sensitivity of the dispersion relation on $\Delta P$, both in the non-dispersive, long-wavelength regime relevant to the determination of the Moens-Korteweg velocity (equation \ref{eq:MK}), and in the dispersive regime relevant to elastography.

We now focus on the impact of pre-stress on the dispersion of axial waves in both regimes. 

\section{Influence of pre-stress}
\label{sec:prestress}
Applying a pressure difference $\Delta P$ leads to an initial out-of-plane deflection $\delta(z)$ (figure \ref{fig1}a).
This static deformation both induces a transversal stretching of the strip, known to affect the dispersion of in-plane modes \cite{delory2024}, and creates curvature, which enhances the static rigidity of shells \cite{reissner1946,Lazarus2012,vella2012, paulsen2016, mahadevan2004}. %by preventing stretch-free deformations 
To disentangle the influence of these two effects on the dispersion relation of the out-of-plane mode, we perform independent measurements, by imposing either an in-plane transversal stretch or a constant curvature.
We then combine these two effects by applying a pressure difference $\Delta P$ using air as a fluid, allowing to neglect the fluid coupling.
This step by step approach empowers us to propose a model that takes into account the influence of both stretching and curvature on waves propagating in pressurized soft conduits.

\subsection{Transversal stretching of the strip}
\label{subsec:stretching}
\begin{figure}[t]
    \centering
    \includegraphics[width = \textwidth]{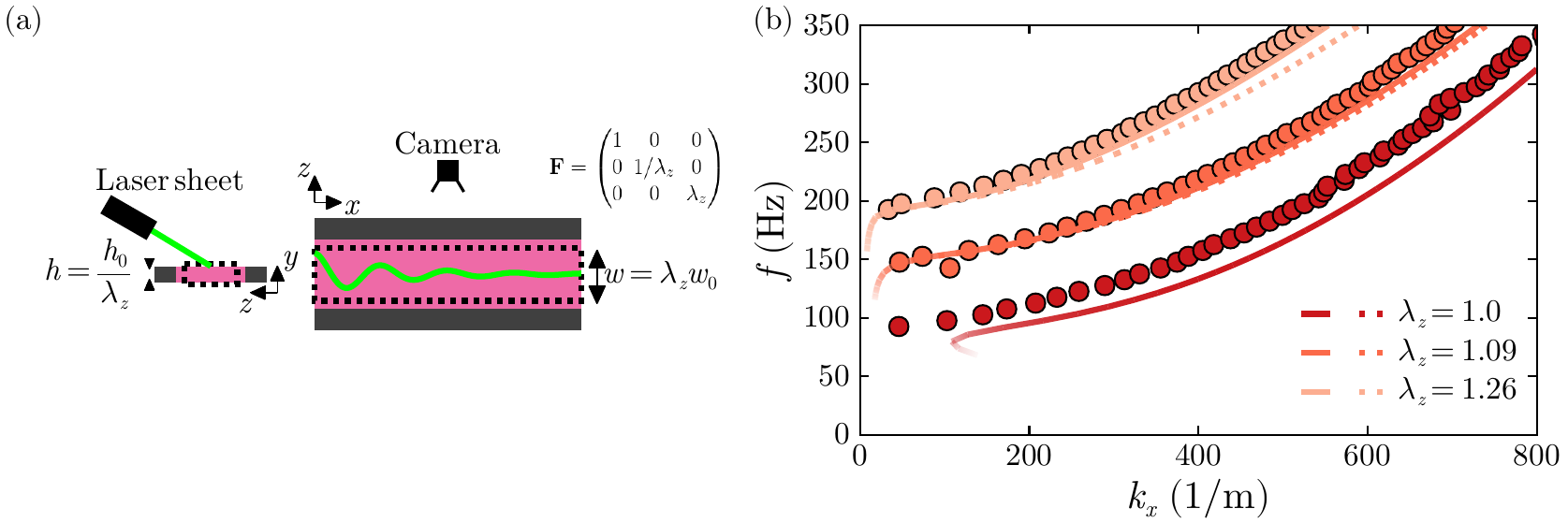}
    \caption{(a) A soft strip, clamped along its edges, is subjected to an elongation in the $\boldsymbol{e}_z$ direction. 
    Elastic waves are generated using a shaker for three imposed values of the stretch ratio $\lambda_z$, and we extract the strip out-of-plane displacement $u_y(x,t)$ by recording the in-plane motion of a laser sheet in oblique incidence. 
    (b) Dispersion curves of the lowest order flexion mode propagating in the strip for $\lambda_z = 1.0$, $1.09$, and $1.26$.
    \revP{The lines are predictions for two distinct hyperelastic models, the Mooney-Rivlin model (solid lines) and the neo-Hookean model (dotted lines), whose transparency renders the ratio $\mathrm{Im}(k_x)/|k_x|$.}
\label{fig3}}
\end{figure}

We clamp the soft elastomer strip (with width $w_0$ and thickness $h_0$) along its long edges, and impose an in-plane transversal stretch by adjusting the distance between the clamps $w = \lambda_z w_0$, where $\lambda_z$ is the transversal stretch ratio (figure \ref{fig3}a).
The boundary conditions prevent axial strain, so that the axial stretch ratio is one and, in the incompressible limit, the strip thickness is reduced by a factor $1/\lambda_z$. We identify a pure shear, homogeneous deformation whose deformation gradient $\boldsymbol{F}$ is represented in figure \ref{fig3}(a).
We then excite small amplitude out-of plane waves in the strip as in section \ref{sec:arteryexperiment}.
This time, as there is no water under the strip, the pattern deformation caused by refraction is minimal and we use a different technique to detect the out-of-plane displacement. A CCD camera records the deflection of a laser sheet projected in oblique incidence on the strip with an acquisition frequency of $\qty{800}{\hertz}$.
The out-of-plane displacement $u_y(x,t)$ is directly proportional to the in-plane deflection of the laser line, which we extract from the images.
We then obtain the dispersion relation of axial out-of-plane waves by performing the two-dimensional Fourier transform of $u_y(x,t)$ and extract maxima in the $(k_x, f)$ plane.

We show in figure \ref{fig3}(b) the dispersion relation for three imposed values of $\lambda_z$ (filled circles). 
Strikingly, we do not observe the non-dispersive low-wavenumber regime evidenced in figure \ref{fig2}(c). 
Indeed, this regime arises from an hybridization between the fundamental mode of the fluid waveguide and the flexion mode of the strip (as detailed in section \ref{sec:fluidcoupling}).
Instead, the dispersion relations have a parabolic shape, and exhibit cutoff frequencies, originating in the clamped boundary condition along the lateral edges.
Imposing a transversal stretch ($\lambda_z > 1$), shifts the dispersion curves upwards, indicating that tension orthogonal to the direction of propagation stiffens the strip. Yet, the propagation remains dispersive, unlike in the string-like regime reached when tension is applied along the principal direction of a beam \cite{delory2024}.
We also observe that the dispersion relation obtained in figure \ref{fig2}(c) when imposing $\Delta P = \qty{0}{\pascal}$ in the water channel is shifted downwards compared to that for $\lambda_z = 1.0$ when there is no channel at all, suggesting that the influence of water coupling partly originates from an added mass effect.

We model the influence of in-plane stretch using the acoustoleastic theory \cite{ogden1997,destrade2007}, which describes incremental motions superimposed on a large deformation.
We replace the elasticity tensor $\boldsymbol{C}$ in equation \eqref{eq:elastodynamics} by the modified elasticity tensor $\boldsymbol{C^0}(\lambda_z, \omega)$, which incorporates both the effects of pre-stress and of viscoelasticity \cite{delory2023}.
We solve this equation, where $\boldsymbol{u}$ now stands for the incremental displacement, with boundary conditions of vanishing incremental traction at the free surfaces ($y = \pm h/2$, where $h =h_0/\lambda_z$), and zero incremental displacement at the clamped edges ($z = \pm w/2$).
We use a solution procedure analogue to that detailed in section \ref{sec:pipewaves} with the displacement form taken as $\boldsymbol{u} = \boldsymbol{u}(k_x, y, z, \omega)e^{\mathrm{i}(k_x x - \omega t)}$, and derivatives discretized in the $y$ and $z$ directions.
Our implementation is identical to that of \cite{kiefer2024_2}, and is available in \cite{chantelot2025}.
The predictions (solid lines) are in satisfying agreement with our measurements (filled circles) for the three values of $\lambda_z$ (figure \ref{fig3}b), suggesting that the modified elasticity tensor successfully captures the influence of the transversal stretch. 
We note that the agreement is not as good for $\lambda_z = 1.0$, revealing the experimental difficulty in ensuring a stress free reference configuration.  
Such variations in the experimental procedure constitute the dominant source of uncertainty in our measurements. 
We repeated experiments (not shown) and observed that the deviation between prediction and measurement in figure \ref{fig3} for $\lambda_z = 1$ is representative of the error in the reference configuration and that this error vanishes as pre-stress is applied.
We encode the ratio of the imaginary part to the absolute value of the wavenumber in the transparency of the prediction lines, showing that stretching facilitates wave propagation at small wavenumbers, in agreement with the increase of the attenuation distance with pressure reported in figure \ref{fig2}(b).
It highlights that accounting for the viscoelastic properties of the soft elastomer is crucial to obtaining quantitative agreement.
\revP{We also evidence the role of material non-linearity by computing the dispersion relations for a different hyperelastic model.
We represent predictions for the neo-Hookean constitutive law, that is the Mooney-Rivlin model with $\alpha = 0$, with dotted lines in figure \ref{fig3}b.
For $\lambda_z = 1.26$, the dispersion relations predicted by the two models deviate at large $k_x$, indicating that material non-linearity influences wave propagation even at moderate elongations, and needs to be taken into account to obtain accurate predictions.}

\subsection{Strip curvature}

\begin{figure}[t]
    \centering
    \includegraphics[width = 0.5 \textwidth]{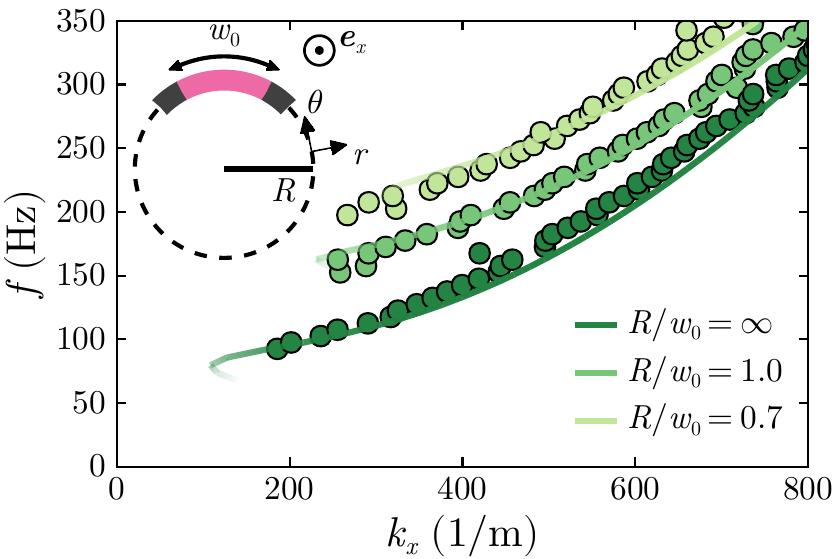}
    \caption{Dispersion curves of the lowest order flexural mode propagating in clamped curved strips with normalized radius of curvature $R/w_0 = \infty$, $1.0$, and $0.7$. The solid lines are predictions obtained without fitting parameters  whose transparency renders the ratio $\mathrm{Im}(k_x)/|k_x|$.}
    \label{fig4}
\end{figure}

We now investigate the influence of curvature on the propagation of out-of-plane waves.
To this end, we design and build supports in order to impose a curvature $R$ to the clamped strip while ensuring that its mid-surface is not stretched, as sketched in the inset of figure \ref{fig4}.
Bending the strip into a cylindrical shape creates both transversal stretching and compression, that is in the direction $\boldsymbol{e}_\theta$ as it is now natural to use cylindrical coordinates.
The maximal value of the stretch ratio in the azimuthal direction can be geometrically estimated as $\lambda_\theta = 1 + h_0/(2R)$ so that, as $h_0 \ll R$, we assume in the following that the curved strip is stress free.
As in section \ref{subsec:stretching}, we use the laser sheet to measure the dispersion relation of axial waves for three imposed radii of curvature $R$ (filled circles in figure \ref{fig4}).
The frequency-wavenumber curves retain their parabolic shape but the cutoff frequency increases as $R$ decreases, and the whole dispersion relation is shifted upwards.
It appears that the curvature-induced rigidity effect identified in static indentation tests \revP{of soft shells} \cite{reissner1946,Lazarus2012,vella2012} \revP{and in wrinkling of curved surfaces \cite{mahadevan2004, paulsen2016}} carries over to the first axial flexion mode.
Qualitatively, this stiffening can be understood using Gauss' \emph{Theorema Egregium} \cite{audoly2010}: the deformation associated to the propagation of axial flexion waves on the curved strip is a non-isometric transformation. It thus generates in-plane stresses which act to rigidify the strip, as evidenced in section \ref{subsec:stretching}. 

We include curvature in our model by solving the elastodynamic equation in a cylindrical geometry, performing the discretization on $r$ and $\theta$, with the same boundary conditions as in section \ref{subsec:stretching} (details of the implementation are given in \cite{chantelot2025}).
We emphasize here that, as the curved strip is assumed to be stress free, the modified elasticity tensor $\boldsymbol{C^0}(\lambda_\theta = 1, \omega)$ only accounts for the rheological behavior of the elastomer.
The computed dispersion relations of the first flexion mode (solid lines in figure \ref{fig4}) are in excellent agreement with our experimental measurements, validating a posteriori the hypothesis of a stress free strip.
%We also observe, both in experiments and numerics, that waves do not propagate below a cutoff wavenumber that increases when decreasing the radius of curvature.

\subsection{Imposed pressure: combining stretching and curvature}
\label{sec:airpressure}

\begin{figure}
    \centering
    \includegraphics[width = 0.75\textwidth]{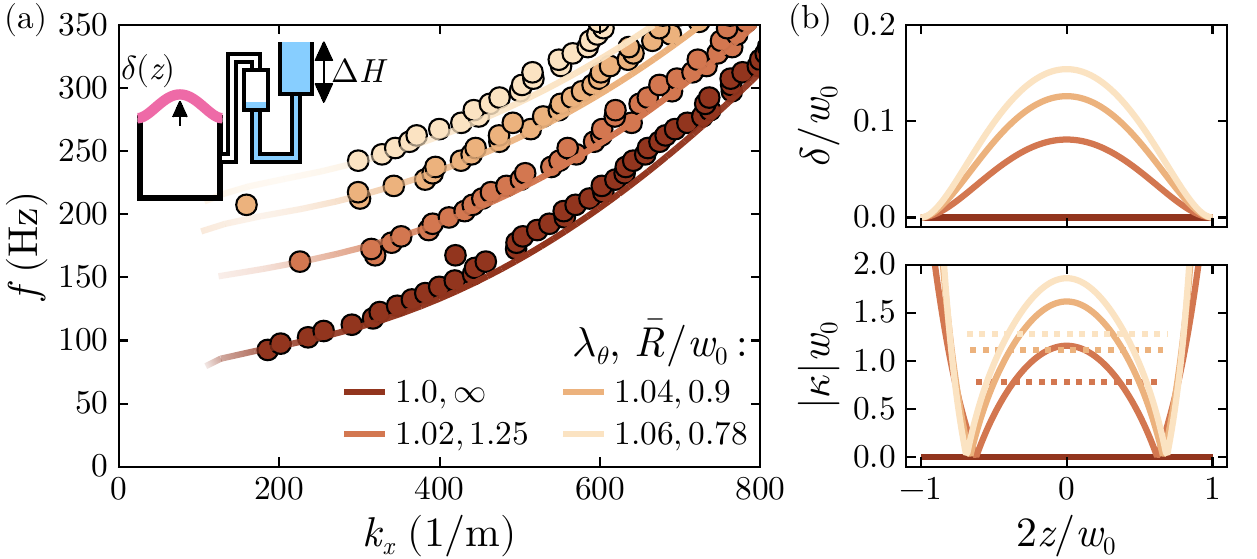}
    \caption{(a) Dispersion relations of the lowest order mode propagating in the model soft pipe for four imposed air pressures $\Delta P = \tikzcircle[fill=Reds1]{2.5pt} \thinspace \qty{0}{\pascal}$, $\tikzcircle[fill=Reds2]{2.5pt} \thinspace \qty{200}{\pascal}$, $\tikzcircle[fill=Reds3]{2.5pt} \thinspace \qty{490}{\pascal}$ and $\tikzcircle[fill=Reds4]{2.5pt} \thinspace \qty{780}{\pascal}$. The solid lines are theoretical predictions, whose transparency encodes the ratio $\mathrm{Im}(k_x)/|k_x|$, for which the input stretch ratio $\lambda_\theta$ and curvature radius $\bar{R}$ are obtained from the static problem.
    (b) Normalized strip deflection $\delta(z)/w_0$ and curvature $|\kappa(z)| w_0$ profiles computed for the same imposed air pressures as in (a).
    }
    \label{fig5}
\end{figure}

Having identified and modeled the respective influence of stretch and curvature on the bending mode, we explore their combined effect when imposing a pressure difference $\Delta P > 0$.
In order to single out the influence of pre-stress from that of fluid coupling, we use air instead of water pressure in the deformable conduit.
We insert a rigid vertical cylinder between the water column and the fluid waveguide (inset of figure \ref{fig5}a).
This chamber allows water to act as a piston, compressing air in the waveguide with a pressure controlled by the height difference between the water in the column and in the intermediate chamber.
In this configuration, we use the same experimental technique as in section \ref{subsec:stretching} to obtain the dispersion relations of axial out-of-plane waves for four values of $\Delta P$, and represent them in figure \ref{fig5}(a).
Unlike for a water filled conduit, the dispersion relations exhibit a parabolic shape and a cutoff frequency, highlighting the need of a strong coupling to observe an hybridization between the dispersion relation of the fluid waveguide and that of the strip.
The curves are shifted upwards when increasing $\Delta P$, as expected from the combined influence of tension and curvature that work together to stiffen the deformable wall.

The first step to model the influence of pressure on the dispersion relation is to link $\Delta P$ to the stretch and curvature experienced by the strip, that is to describe the static finite deformation.
We obtain the deflection $\delta(z)$ \revP{using the Föppl-von K\'arm\'an equations that describe large deflections of thin plates at leading order when considering both geometrical and material non-linearities \cite{erbay1997,dervaux2009}.}
\revP{A detailed justification of our approach, and the} solution to the static problem are provided in appendix \ref{app:B}.
We show in figure \ref{fig5}(b) the normalized midline deflection profiles, $\delta(z)/w_0$, corresponding to the values of $\Delta P$ imposed in the experiments of figure \ref{fig5}(a).
From these profiles, we can determine the membrane stretch ratio $\lambda = \ell / w_0$, where $\ell$ is the length of the deformed strip mid-surface, and obtain the normalized curvature profiles $|\kappa(z)|w_0$ (right panel of figure \ref{fig5}b).
The imposed pressure generates a non-constant curvature, which exhibits strong variations near the edges of the strip because of the clamped boundary condition. 

Accounting for the intricate details of this complex deformed configuration would require finite element simulations.
Instead, we propose to introduce a simpler equivalent geometry, in which we evidence how pre-stress affects the propagation of the first flexion mode.
We consider a clamped cylindrical strip whose radius of curvature and azimuthal stretching depend on $\Delta P$.
We take the radius of curvature, denoted $\bar{R}$, as the average of the curvature profile obtained from the static problem in the central region (dashed lines in the right panel of figure \ref{fig5}b), and assume that the equivalent strip is homogeneously stretched in the azimuthal direction with stretch ratio $\lambda_\theta = \lambda$.
In this simplified geometry, the deformation is homogeneous and we can solve the elastodynamics equation \eqref{eq:elastodynamics} with the modified elasticity tensor $\boldsymbol{C^0}(\lambda_\theta = \lambda, \omega)$, with boundary conditions of vanishing incremental normal traction, and vanishing incremental displacement at $r = \bar{R} \pm h/2$ and $\theta = \pm w/(2\bar{R})$, respectively.
The dispersion relations of the lowest order out-of plane modes computed for the couples ($\lambda_\theta$, $\bar{R}$) determined from the static problem are displayed as solid lines in figure \ref{fig5}(a).
Despite our strong modeling assumptions, the predictions are in excellent agreement with the experimental data (filled circles). 
The model quantitatively captures the real part of $k_x$, and we also observe a qualitative agreement for its imaginary part: waves are not detected when we expect a large attenuation (encoded in the transparency of the solid lines).
We emphasize that, in the parameter range considered in this study, the observed dispersion depends significantly on both $\bar{R}$ and $\lambda_\theta$, as expected from scaling relations obtained for the indentation of pressurized soft shells \cite{Lazarus2012}. 
\revP{The interplay between curvature and in-plane stretching is complex \cite{vella2012, rallabandi2019}: both influences cannot be disentangled by inspecting the dispersion relation.}

We have evidenced that transverse initial stretch and strip curvature both speed up the propagation of out-of-plane elastic waves.
When applying a static pressure difference in the absence of water, these two effects combine through the deflection of strip, and we capture their interplay in an equivalent geometry that enables us to retrieve the propagation properties of the strip's first flexion mode.
In the following, we investigate how the soft, pre-stressed strip couples to the fluid waveguide in the presence of water.

\section{Influence of water loading}
\label{sec:fluidcoupling}
Flexion waves propagating in the soft strip set the surrounding water in motion, thereby increasing the inertia of the coupled system.
This added mass effect decreases the wave speed, as observed when comparing the experiments of sections \ref{sec:arteryexperiment} and \ref{sec:prestress} (figures \ref{fig2}c and \ref{fig5}a).
The volume of fluid set in motion can be described using a penetration length. Indeed, as flexural waves propagate much slower in the soft strip than pressure waves in the fluid ($V_f = \qty{1500}{\meter\per\second}$), we expect evanescent waves in the surrounding water.
We distinguish two asymptotic regimes: (i) a short-wavelength regime when the penetration length is smaller than the waveguide depth $w_0$, and (ii) a long-wavelength regime when the penetration length is larger than $w_0$, where the finite size of the fluid conduit must be taken into account.

\subsection{Short-wavelength regime}
\begin{figure}
    \centering
    \includegraphics[width = \textwidth]{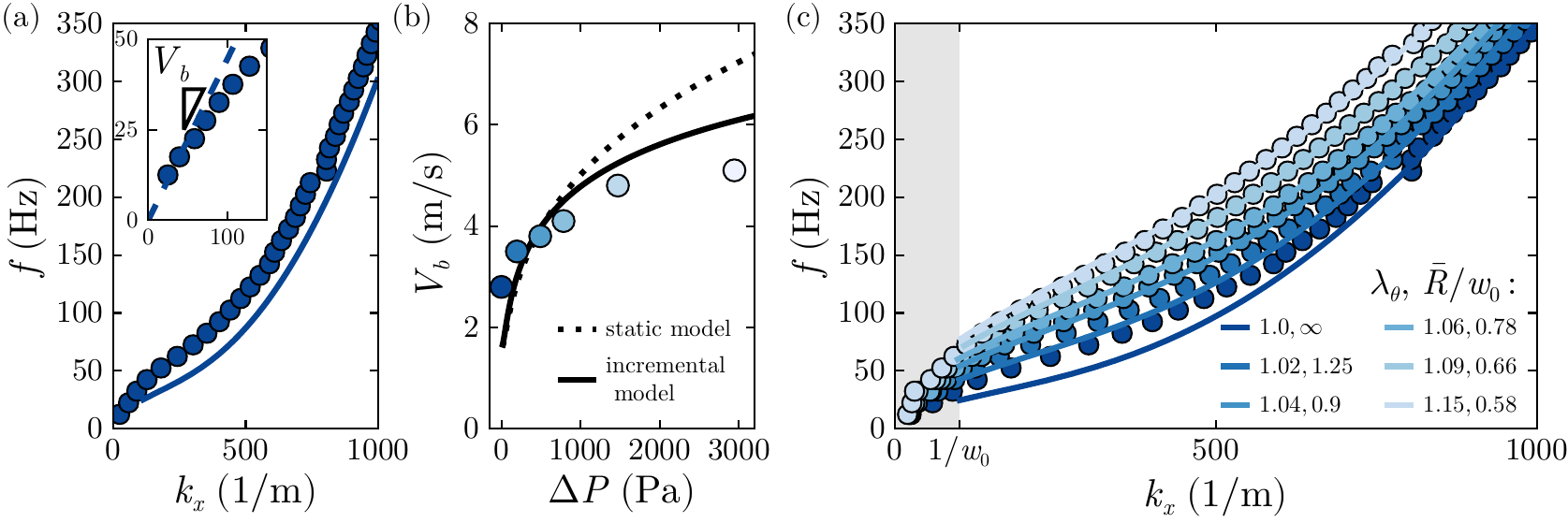}
    \caption{(a) Dispersion relation in the water filled soft pipe for $\Delta P = \qty{0}{\pascal}$. 
    In the long-wavelength limit, the propagation is not dispersive as evidenced by the dashed line in the inset, while for $k_xw_0 \gg 1$, the dispersion is similar to that predicted for a strip in contact with a semi-infinite fluid (solid line).
    (b) Velocity of the breathing mode $V_{b}$ measured using a time of flight technique as a function of $\Delta P = \tikzcircle[fill=Blues1]{2.5pt} \thinspace \qty{0}{\pascal}$, $\tikzcircle[fill=Blues2]{2.5pt} \thinspace \qty{200}{\pascal}$, $\tikzcircle[fill=Blues3]{2.5pt} \thinspace \qty{490}{\pascal}$, $\tikzcircle[fill=Blues4]{2.5pt} \thinspace \qty{780}{\pascal}$, $\tikzcircle[fill=Blues5]{2.5pt} \thinspace \qty{1500}{\pascal}$, and $\tikzcircle[fill=Blues6]{2.5pt} \thinspace \qty{2900}{\pascal}$.
    (c) Dispersion relations in the water filled model soft pipe for all $\Delta P$. 
    The solid lines are predictions for the short-wavelength regime $k_x w_0 \gg 1$, which are fitted to the experimental data by multiplying all wavenumbers by a factor 0.93.}
    \label{fig6}
\end{figure}

We first consider the $k_x w_0 \gg 1$ regime in which the dispersion relation is similar to that of the first flexion mode of the strip, as evidenced in figure \ref{fig6}(a) for the water filled conduit with $\Delta P = \qty{0}{\pascal}$.
In this regime, we assume that the strip is coupled to a semi-infinite water medium.
Similarly as in section \ref{sec:pipewaves}, we model the presence of the surrounding fluid as a boundary condition to the solid domain.
Unlike in the case of a plate or a cylinder \cite{gravenkamp2014}, there is no general expression for the normal stress, leading us to use an approximate boundary condition to obtain the dispersion relation of the first flexion mode.
We assume that the solution in the water domain corresponds to a plane wave, with wavevector $\boldsymbol{k}_f$. 
The acoustic pressure at the water-strip interface reads
\begin{equation}
    P'(k_x,y = -h/2,z,\omega) = \rho_f \omega^2 \frac{\boldsymbol{u}\cdot\boldsymbol{e}_y}{\mathrm{i}\boldsymbol{k}_f\cdot\boldsymbol{e}_y},
    \label{eq:fluidp}
\end{equation}
where we invoked the continuity of the normal displacement at the interface \cite{kiefer2019}.
We use Snell's law to express the normal projection of the fluid wavevector at the interface $(\boldsymbol{k}_f\cdot\boldsymbol{e}_y)^2 = \omega^2/V_f^2 - k_x^2 - k_z^2$, and further simplify this expression to $(\boldsymbol{k}_f\cdot\boldsymbol{e}_y)^2 \approx  - k_x^2$ as (i) for the first flexion mode in the short-wavelength limit $k_z^2 \ll k_x^2$, and (ii) the longitudinal wave velocity in the fluid is much larger than the speed of the flexion mode. 
This approximation evidences the presence of evanescent waves and equation \eqref{eq:fluidp} thus underlines that the surrounding fluid acts as an added mass as $P' \propto \rho_f \omega^2$.
%We remark that this approximation corresponds to the presence of evanescent waves in the fluid, and that equation \eqref{eq:fluidp} underlines that the surrounding fluid acts as an added mass as $P' \propto \rho_f \omega^2$.
We implement this approximate boundary condition and represent the predicted dispersion relation for $\Delta P = \qty{0}{\pascal}$ in figure \ref{fig6}(a) (solid line). 
Although the prediction satisfactorily captures the experimental dispersion relation shape for short-wavelengths (in the high $k_x$ region), it is shifted downwards compared to the experimental data, suggesting that we overestimate the water loading.
We attribute this mismatch to edge effects: we do not model the presence of the side walls of the fluid waveguide which would reduce the volume of fluid set in motion.

We further test the relevance of this approximate boundary condition in the presence of pre-stress.
We compute predictions for each $\Delta P$ using the equivalent geometry introduced in section \ref{sec:airpressure}, that of a curved, azimuthally stretched clamped strip, where $\lambda_\theta$ and $\bar{R}$ are extracted from the static problem (solid lines in figure \ref{fig6}c).
We find a good agreement between the experimental data and the short-wavelength predictions for all imposed $\Delta P$ by systematically multiplying the calculated $k_x$ by a factor $0.93$, validating the use of the approximate boundary condition.

\subsection{Long-wavelength regime}
In the long-wavelength limit, we observe the most remarkable feature of water loading.
The finite size of the fluid domain leads to the hybridization between the fundamental mode of the fluid waveguide and the flexion mode of the strip, giving rise to a non dispersive behavior (inset of figure \ref{fig6}a).
We highlight that, in this limit, waves propagate at a speed of a few meters per second, \emph{i.e.} three orders of magnitude slower than pressure waves in a rigid waveguide.
While the experiments of section \ref{sec:arteryexperiment} (figure \ref{fig2}a) enable us to precisely measure the dispersion relation for $k_x w_0 >1$, the inset of figure \ref{fig6}(a) evidences that the size of the experimental setup limits the resolution in $k_x$, preventing reliable measurements of the breathing mode velocity $V_b$.
We thus perform complementary measurements to determine $V_b$ using a time-of-flight approach.
To increase the signal to noise ratio, we excite small amplitude waves with a one second long quadratic chirp signal with instantaneous frequency ranging from $\qty{1}{\hertz}$ to $\qty{100}{\hertz}$, and obtain the impulse response by correlation of the measured displacement field $u_y(x,z,t)$ with the time reversed input signal.
This response is then low-pass filtered, using a Butterworth filter with cutoff frequency of $\qty{6}{\hertz}$, to single out the non-dispersive regime.
The wave speed is the slope of the propagating wavefront visualized in the spatiotemporal representation of the filtered impulse response.
Figure \ref{fig6}(b) displays $V_b$ for all imposed $\Delta P$, where the slope angle is determined as the maximum of the Radon transform of the spatiotemporal $(x, t)$ map. The wave velocity significantly increases with $\Delta P$, underlining the influence of pre-stress in the long-wavelength regime.

We follow the same approach as when deriving the Moens-Korteweg velocity to obtain a prediction for $V_b$. 
Performing mass and momentum balances in the fluid contained in the soft conduit yields the speed of a pressure disturbance, $V_b = \sqrt{A/\left(\rho_f\mathrm{d}A/\mathrm{d}\Delta P\right)}$, where $A$ is the cross-section of the fluid conduit.
To estimate $\mathrm{d}A/\mathrm{d}\Delta P$, it is tempting to use the results of the static deflection problem by numerically integrating  $\delta(z)$ along the conduit width, yielding the area $A(\Delta P)$ which we can then differentiate to get its variation with imposed pressure.
The prediction obtained following this approach is represented with a dotted line in figure \ref{fig6}(b).
While it qualitatively captures the increasing trend of $V_b$ with the imposed pressure, it systematically and significantly overestimates the wave speed at large $\Delta P$.

We address this shortcoming by introducing the following incremental approach, which recognizes that we excite small amplitude waves on top of a finite deformation.
We consider the equivalent geometry of section \ref{sec:airpressure}. 
Our goal is to obtain the deflection $\delta'(z)$ of the equivalent cylindrical clamped strip, with azimuthal stretch $\lambda(\Delta P)$ and radius of curvature $\bar{R}(\Delta P)$, upon applying an incremental pressure load $P'$.
We derive an expression for $\delta'(z)$ using shallow shell theory \cite{ventsel2001}, taking into account the background strip tension and the strip curvature but leaving out material non-linearities, as detailed in appendix \ref{app:C}.
We are thus able to numerically estimate the area increase in the deformed configuration $A' = \int \delta' \mathrm{d}z$, and its variation with $P'$.
In this incremental approach, the velocity $V_b$ is given by,
\begin{equation}
    V_b = \lim_{P'\to 0} \left(\sqrt{\frac{A}{\rho_f \mathrm{d}A'/\mathrm{d}P'}}\right).
    \label{eq:vbincr}
\end{equation}
This prediction, represented with a solid line in figure \ref{fig6}(b), is more consistent with our data, underlining the need to distinguish the reference and deformed configurations for large deformations. 
Yet, we still overestimate $V_b$ at large $\Delta P$, a discrepancy that could be attributed to our \revP{chosen simplified geometry and our} neglect of the material non-linearities.

\section{Conclusion}
\label{sec:discussion}
We combine theory and experiments to investigate the propagation of guided waves in a model artery with the aim to assess the link between the wave speed and the properties of the arterial wall.  
We first obtain numerically the dispersion relation of low frequency guided waves propagating in soft elastic pipes filled and surrounded by water, and focus on the fundamental breathing and flexion modes that have been observed \emph{in vivo} \cite{laloy2023}.
We identify two key pieces of information relevant to medical diagnosis.
(i) In the frequency range of interest for shear wave elastography, both modes are dispersive, indicating that wave guiding must be taken into account to infer the wall elasticity, in agreement with literature for the breathing mode.
(ii) The flexural pulse wave velocity is highly sensitive to the pipe's surrounding, an effect we quantitatively capture in a long-wavelength model, making it less suitable than the breathing pulse wave to determine the arterial wall properties.

We then build a minimal artery phantom which reproduces the physics of the breathing mode. 
We explore the influence of pre-stress by varying the internal pressure, and observe that out-of-plane waves are sensitive to the imposed pressure both in the long-wavelength regime relevant to pulse wave propagation, and in the short-wavelength regime relevant to shear wave elastography.
By singling out their contribution in dedicated experiments, we show how transversal stretching and curvature-induced rigidity combine to stiffen the pressurized phantom through the deflection of the soft wall.
This approach allows us to develop a semi-analytical model, based on the acoustoelastic theory, that quantitatively captures the influence of imposed pressure on the breathing mode.

Finally, we address the coupling with the fluid waveguide by identifying the added mass associated to fluid motion in the limiting cases of small and large wavelengths. 
We draw a parallel between the low-wavenumber non-dispersive behavior in our experiments, stemming from the hybridization between fluid pressure waves and the first flexion mode of the clamped strip, and the pulse wave velocity.
This enables us to show how the Moens-Korteweg formula can be reframed using an incremental approach in the presence of pre-stress to capture the increase of the low-wavenumber breathing wave velocity $V_b$ with $\Delta P$.

We now discuss how the observations made in our model experiment carry over to arteries.
First, pressurisation creates transversal stretching and curvature in the phantom.
In contrast, imposing a pressure difference in a pipe causes stretching but reduces curvature.
The influence of these different behaviors on the dispersion relation of the fundamental breathing mode is best understood in the long-wavelength limit.
In the phantom, curvature-induced rigidity leads to an increase of $V_b$ with $\Delta P$.
For a pipe, the Moens-Korteweg formula, \eqref{eq:MK}, predicts the opposite trend, through an increase of tube radius and a decrease of wall thickness.
Reconciling the Moens-Korteweg prediction with the blood pressure induced pulse wave acceleration measured \emph{in vivo} \cite{histand1973, marais2018, baranger2023} highlights a second difference between real and model arteries.
Indeed, the increase of $V_b$ with $\Delta P$ is attributed to the complex structure of the arterial tissue which leads to strongly non-linear stiffening at physiological blood pressures: the Moens-Korteweg formula involves the incremental Young modulus \cite{pedley1980}.
Our experiments thus reproduce the \emph{in vivo} behavior but not its underlying cause
as material non-linearities dominate \emph{in vivo} while geometrical non-linearities prevail in the phantom.

To conclude, we evidence that quantitative predictions of wave propagation in the model artery require to take into account wave guiding, geometrical non-linearities, material non-linearities and their interplay with viscoelastic material properties. 
These results emphasize that pulse wave velocity and elastography measurements should always be analyzed in light of the applied pre-stress, and pave the way for imaging techniques that go beyond the determination of incremental elastic moduli.
These findings also suggest to investigate and model the influence of other forms of pre-stress acting on arteries, such as axial extension or stresses induced by tissue growth, and to study materials that have more pronounced non-linear material properties upon pressurisation and/or a layered architecture to better mimic the arterial wall.
Going further, investigating the combined influence of pressure loading and axial stretching on the propagation of elastic waves in soft tubes could enable to predict the occurrence of bulging bifurcations, that is the possible formation of aneurysms \cite{haughton1979a,guo2022}.
Finally, we expect blood flow to influence the propagation of arterial waves \cite{histand1973}, an effect that the experimental setup developed in this work will enable us to probe.

%% Final declarations after the Conclusion

\appendix
\section{Mooney-Rivlin model and modified viscoacoustoelastic tensor}
\label{app:A}
We describe the behavior of Ecoflex OO-30 using the compressible Mooney-Rivlin model. 
The hyperelastic constitutive law relies on the strain energy density function $W$ which is built on the principal invariants of the left Cauchy-Green tensor $\boldsymbol{B} = \boldsymbol{F}\cdot\boldsymbol{F}^T$,
\begin{equation}
    W = \frac{\mu_0}{2} \left[\left(1-\alpha\right)\left(\frac{I_1}{J^{2/3}}-3\right) + \alpha \left(\frac{I_2}{J^{4/3}}-3\right)\right] + \frac{\kappa}{2}\left(J-1\right)^2,
\end{equation}
with, 
\begin{align*}
    &I_1 = tr(\boldsymbol{B}) = \lambda_1^2 + \lambda_2^2 +\lambda_3^2, \\
    &I_2 = \frac{1}{2}\left(\mathrm{tr}(\boldsymbol{B})^2 - \mathrm{tr}(\boldsymbol{B}^2)\right) = \lambda_2^2\lambda_3^2 + \lambda_1^2\lambda_3^2 + \lambda_1^2\lambda_2^2, \\
    &I_3 = \mathrm{det}(\boldsymbol{B}) = \lambda_1^2\lambda_2^2\lambda_3^2 = J^2, \\
    &\kappa = \rho_s V_L^2.
\end{align*}
From $W$, an incremental approach allows to build a modified elasticity tensor $\boldsymbol{C^0}(\lambda_1,\lambda_2,\lambda_3)$ which describes the material response to small displacements superposed on the large deformation described by $\boldsymbol{F}$ \cite{ogden1997,destrade2007}.
The coefficients of $\boldsymbol{C^0}$ are given in \cite{delory2024} and are identical to that given in \cite{ogden1997,destrade2007} with a permutation of the last two indices.

The effect of viscoelasticity is also incorporated in the modified elasticity tensor $\boldsymbol{C^0}$ which now becomes frequency dependent and reads,
\begin{equation}
    C^0_{jikl}(\lambda_1,\lambda_2,\lambda_3, \omega) = C^0_{jikl}(\lambda_1,\lambda_2,\lambda_3) 
    +\mu_0I_{jikl}\left(1 + \beta'\frac{\lambda_i^2+\lambda_j^2-2}{2}\right)(\mathrm{i}\omega\tau)^n,
\end{equation}
with $I_{jikl} = (\delta_{jk}\delta_{il}+\delta_{jl}\delta_{ik})$, and $\beta' = 0.29$ for our soft elastomer \cite{delory2023}.
The second term, that involves both the principal stretches and the frequency, underlines the coupling between pre-stress and viscoelasticity.

\section{Static deflection}
\label{app:B}
We consider the static deflection $\delta$ of a rectangular plate with a large aspect ratio clamped along its long edges and loaded with an imposed pressure $\Delta P$, as sketched in figure \ref{fig1}(a).  
We look for solutions that are invariant along the $x$ direction under the assumptions that the plate is thin ($h_0 \ll w_0$), that the deflection can be large ($\delta > h_0$), and that the strain remains small ($\delta \ll w_0$).
With these hypothesis, the deflection $\delta(z)$ obeys the Föppl-von K\'arm\'an equation \cite{landau1970},
\begin{equation} 
    \frac{E_0 h_0^3}{12\left(1-\nu^2\right)}\frac{\mathrm{d}^4\delta}{\mathrm{d}z^4} - h_0\frac{\partial}{\partial x_i} \left(\sigma_{zi} \frac{\mathrm{d}\delta}{\mathrm{d}z} \right) = \Delta P,
    \label{eq:FvK0}
\end{equation}
with clamped boundary conditions,
\begin{align}
    \begin{split}
        &\delta(z = \pm w_0/2) = 0 \\
        &\left.\frac{\mathrm{d}\delta}{\mathrm{d}z}\right\vert_{z = \pm w_0/2} = 0,
    \end{split}
    \label{eq:FvKbc}
\end{align}
where $\nu$ is the Poisson ratio and $\boldsymbol{\sigma}$ is the Cauchy stress tensor.
\revP{Equation \eqref{eq:FvK0} gives the leading order description of the static problem in the presence of both geometrical and material non-linearity \cite{erbay1997, dervaux2009}.}
It can be simplified by realizing that the stresses $\sigma_{zy}$, and $\sigma_{zz}$ do not vary significantly across the small thickness of the strip.
From the boundary conditions, we obtain $\sigma_{zy} = 0$ and $\sigma_{zz} = T/h_0$, where $T$ is the unknown membrane tension, so that \eqref{eq:FvK0} becomes,
\begin{equation} 
    \frac{E_0 h_0^3}{12\left(1-\nu^2\right)}\frac{\mathrm{d}^4\delta}{\mathrm{d}z^4} - T \frac{\mathrm{d}^2\delta}{\mathrm{d}z^2} = \Delta P.
    \label{eq:FvK}
\end{equation}
\revP{We note that while the geometrical non-linearity directly appears through the tension term, that implicitly depends on the deflection $\delta$, the material non-linearity is absent in the leading order approximation.}
\revP{This observation is consistent with the acoustoelastic predictions obtained in figure \ref{fig3}(b) where the neo-Hookean and Mooney-Rivlin hyperelastic models give identical results in the small wavenumber limit.
Material non-linearities can be neglected to obtain long-wavelength approximate solutions, such as that of the static problem, but are relevant to accurately predict wave propagation at larger wavenumbers.}
\begin{figure}[t]
    \centering
    \includegraphics[width = 0.66\textwidth]{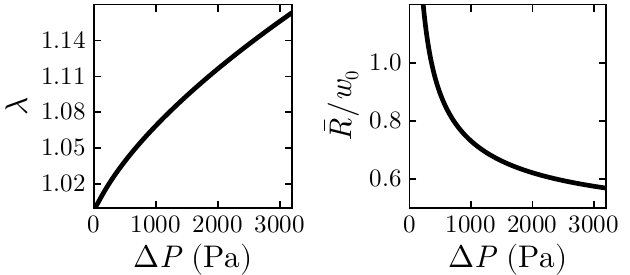}
    \caption{Membrane stretch ratio $\lambda = \ell / w_0$ and normalized average central curvature $\bar{R} / w_0$ as a function of $\Delta P$.}
    \label{figB}
\end{figure}

We determine the static deflection $\delta(z)$ as a solution of equation \eqref{eq:FvK} with boundary conditions given by \eqref{eq:FvKbc}.
The solution reads,
\begin{equation}
    \delta = \frac{\Delta P w_0^2}{8T}\left[1- \bar{z}^2 +\frac{2}{a}\left(\frac{\cosh(a \bar{z})}{\sinh(a)} - \coth(a)\right)\right],
    \label{eq:deflection}
\end{equation}
where $a = w_0\sqrt{T/B}/2$, $B = E_0 h_0^3 /(12(1-\nu^2))$, and $\bar{z} = 2z/w_0 $.
To obtain a prediction for $\delta$, the next step is to elucidate the implicit dependence on the tension.
This is achieved by computing the length of the deformed strip,
\begin{equation}
    \ell = \frac{w_0}{2} \int_{-1}^1 \sqrt{1+\left(\frac{\mathrm{d}\delta}{\mathrm{d}z}\right)^2} \mathrm{d}\bar{z},
    \label{eq:stretch}
\end{equation}
and by relating it to the tension using Hooke's law under a plane strain assumption,
\begin{equation}
    T = \frac{E_0 h_0}{1-\nu^2} \left(\frac{\ell}{w_0} - 1\right).
    \label{eq:tensionlinearelasticity}
\end{equation}

Combining equations \eqref{eq:deflection}, \eqref{eq:stretch}, and \eqref{eq:tensionlinearelasticity} yields an equation that depends only on $T$ and $\Delta P$, enabling us to find $T$ numerically when imposing $\Delta P$.
Having removed the implicit tension dependence, we proceed to get the deflection profile $\delta(z)$ from equation \eqref{eq:deflection}, and the two quantities of interest for modelling wave propagation in the pressurized waveguide the membrane stretch ratio $\lambda = \ell / w_0$ and the normalized average central curvature $\bar{R} / w_0$.
We represent $\lambda$ and $\bar{R} / w_0$ as a function of $\Delta P$ in figure \ref{figB}.

\section{Equivalent strip model}
\label{app:C}
\begin{figure}[t]
    \centering
    \includegraphics[width = 0.66\textwidth]{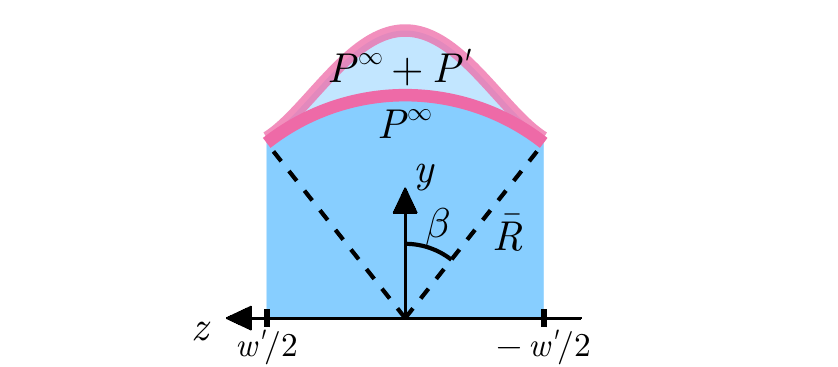}
    \caption{Sketch of the equivalent strip geometry.}
    \label{figC}
\end{figure}

We consider a cylindrical strip with radius of curvature $\bar{R}$, angular opening $2\beta = \lambda w_0 / \bar{R}$, and thickness $h = h_0/\lambda$, where $\lambda$ and $\bar{R}$ are obtained from the static deflection problem, as sketched in figure \ref{figC}.
We model the deformation of this strip using the theory of shallow shells \cite{ventsel2001}.
We assume that the pre-stress in the strip is generated by a background pressure $P^\infty$, and are interested in the deformation when an incremental pressure $P'$ is applied on top of the static load.
We thus use a perturbative approach following \cite{berry1958,vella2012_2}.
The equations of shallow shell theory in cartesian coordinates read,
\begin{align}
    \frac{E_0h^3}{12(1-\nu^2)}\frac{\mathrm{d}^4\delta}{\mathrm{d}z^4} + \frac{1}{\bar{R}}\frac{\partial^2 \phi}{\partial x^2} - \frac{\partial^2 \phi}{\partial x^2}\frac{\mathrm{d}^2\delta}{\mathrm{d}z^2} &= P^\infty +P' \label{eq:FvKRforce} \\
    \frac{\partial^4 \phi}{\partial z^4} + \frac{\partial^4 \phi}{\partial x^4} + 2 \frac{\partial^4 \phi}{\partial x^2 \partial z^2}&= 0, \label{eq:FvKRstrain}
\end{align}
where $\phi$ is the Airy stress function which describes the middle surface stresses, and $\delta(z)$ is the displacement of the shell midline from its initial position $y \approx \bar{R} - z^2/(2\bar{R})$. 
The term $\partial^2_x \phi/\bar{R}$ evidences the coupling between curvature and in-plane stretching responsible for the shell's geometric rigidity.

We first discuss the pre-stress generated by the static load.
The imposed pressure induces membrane stresses,
\begin{equation}
    T^\infty_{zz} = \frac{\partial^2\phi^\infty}{\partial x^2}, \; T^\infty_{xx} = \frac{\partial^2\phi^\infty}{\partial z^2}, \; \mathrm{and} \; T^\infty_{xz} = -\frac{\partial^2\phi^\infty}{\partial x \partial z},
\end{equation}
where the Airy stress function $\phi^\infty$ is assumed to satisfy the biharmonic equation \eqref{eq:FvKRstrain} \cite{berry1958}.
We follow the same approach as in section \ref{sec:airpressure} and hypothesize that the equivalent circular strip is azimuthally stretched by $\lambda$, which amounts to setting the tension as $T^\infty_{zz} = 4/3 E_0 h (\lambda - 1)$, and that the pre-stressed shape remains circular, that is $\delta^\infty$ is a constant.
In doing so, we recover Laplace law from equation \eqref{eq:FvKRforce}, $T^\infty_{zz} = P^\infty \bar{R}$.

We now perturb this pre-stressed configuration by letting $\delta = \delta^\infty + \delta'$ and $\phi = \phi^\infty + \phi'$.
At leading order, equation \eqref{eq:FvKRforce} becomes,
\begin{equation}
    \frac{E_0h^3}{12(1-\nu^2)}\frac{\mathrm{d}^4\delta'}{\mathrm{d} z^4} + \frac{T'_{zz}}{\bar{R}} - T^\infty_{zz}\frac{\mathrm{d}^2\delta'}{\mathrm{d}z^2} = P',
\end{equation}
where $T'_{zz} = \partial^2\phi'/\partial x^2$ is an incremental tension.
The solution to this equation with clamped boundary conditions for the incremental displacement at $z = \pm w'/2 = \pm \bar{R}\sin(\beta)$ is given by equation \eqref{eq:deflection} by letting $T \rightarrow T^\infty_{zz}$, $\Delta P \rightarrow P' - T'_{zz}/ \bar{R}$, and $w_0 \rightarrow w'$.
Similarly as above, we elucidate the implicit dependence on the tension $T'_{zz}$ by computing the length of the deformed shell midline and relating it to the tension using Hooke's law under a plane strain assumption, enabling us to find $T'_{zz}$, and thus $\delta'(z)$ when imposing $P'$.
We then integrate $\delta'$ between $z = -w'/2$ and $z=w'/2$ when systematically varying $P'$, allowing us to numerically estimate the derivative $\mathrm{d} A'/\mathrm{d}P'$.
As we use an incremental approach, the velocity $V_b$ is given by equation \eqref{eq:vbincr}.

We repeat this process for all couples $(\lambda, \bar{R})$ obtained from the static deflection problem, enabling us to compute $V_b$ as a function of the imposed pressure $\Delta P$, and thus to obtain a prediction for the wave velocity in the long-wavelength regime.

\section*{Acknowledgements}
We thank D. A. Kiefer for his help in implementing the spectral collocation method. 

\section*{Conflicts of interest}

The authors do not work for, advise, own shares in, or receive funds from any organization
that could benefit from this article, and have declared no affiliations other than their research
organizations.

\section*{Supplementary materials}
The scripts to compute dispersion relations and data underlying this work are available on Zenodo (DOI: \href{https://zenodo.org/records/16037558}{10.5281/zenodo.16037557}) and \href{https://github.com/pchantelot/Wave_propagation_in_a_model_artery_data}{GitHub}, respectively.

\printbibliography

\end{document}